\documentclass[twocolumn]{aastex631}

\usepackage{booktabs}

\usepackage{graphicx}

\begin{document}

\title{Detection of Carbon Monoxide in the Atmosphere of WASP-39b Applying Standard Cross-Correlation Techniques to JWST NIRSpec G395H Data}

\correspondingauthor{Emma Esparza-Borges}\email{emma.esparza.borges@iac.es}

\author[0000-0002-2341-3233]{Emma Esparza-Borges}\affiliation{Instituto de Astrof\'isica de Canarias, E-38200 La Laguna, Tenerife, Spain}\affiliation{Departamento de Astrof\'isica, Universidad de La Laguna, E-38206 La Laguna, Tenerife, Spain}

\author[0000-0003-3204-8183]{Mercedes L\'opez-Morales} \affiliation{Center for Astrophysics ${\rm \mid}$ Harvard {\rm \&} Smithsonian, 60 Garden St, Cambridge, MA 02138, USA}

\author[0000-0002-4489-3168]{J\'ea I. Adams Redai} \affiliation{Center for Astrophysics ${\rm \mid}$ Harvard {\rm \&} Smithsonian, 60 Garden St, Cambridge, MA 02138, USA}

\author[0000-0003-0987-1593]{Enric Pallé} \affiliation{Instituto de Astrof\'isica de Canarias, E-38200 La Laguna, Tenerife, Spain}\affiliation{Departamento de Astrof\'isica, Universidad de La Laguna, E-38206 La Laguna, Tenerife, Spain}

\author[0000-0002-4207-6615]{James Kirk} \affiliation{Department of Physics, Imperial College London, Prince Consort Road, London, SW7 2AZ, UK}

\author[0000-0002-2891-8222]{Núria Casasayas-Barris} \affiliation{Instituto de Astrof\'isica de Canarias, E-38200 La Laguna, Tenerife, Spain}

\author[0000-0003-1240-6844]{Natasha E. Batalha} \affiliation{ NASA Ames Research Center, Moffett Field, CA 94035, USA}

\author[0000-0002-3627-1676]{Benjamin V.\ Rackham} \altaffiliation{51 Pegasi b Fellow}  \affiliation{Department of Earth, Atmospheric and Planetary Sciences, Massachusetts Institute of Technology, 77 Massachusetts Avenue, Cambridge, MA 02139, USA}  \affiliation{Kavli Institute for Astrophysics and Space Research, Massachusetts Institute of Technology, Cambridge, MA 02139, USA}

\author[0000-0003-4733-6532]{Jacob L.\ Bean} \affiliation{Department of Astronomy \& Astrophysics, University of Chicago, Chicago, IL 60637, USA}

\author[0000-0003-2478-0120]{S.~L. Casewell} \affiliation{Centre for Exoplanet Research, School of Physics and Astronomy, University of Leicester, University Road, Leicester, LE1 7RH, UK}

\author[0000-0002-5342-8612]{Leen Decin}\affiliation{Institute of Astronomy, KU Leuven, Celestijnenlaan 200D, 3001 Heverlee, Belgium}

\author[0000-0002-2248-3838]{Leonardo A. Dos Santos} \affiliation{Space Telescope Science Institute, 3700 San Martin Drive, Baltimore, MD 21218, USA}

\author[0000-0003-1756-4825]{Antonio Garc\'ia Mu\~noz} \affiliation{Universit\'e Paris-Saclay, Universit\'e Paris Cit\'e, CEA, CNRS, AIM, 91191, Gif-sur-Yvette, France}

\author[0000-0002-8955-8531]{Joseph Harrington} \affiliation{Planetary Sciences Group, Department of Physics and Florida Space Institute, University of Central Florida, Orlando, FL, USA}

\author[0000-0003-1907-5910]{Kevin Heng} \affiliation{Ludwig Maximilian University of Munich, Faculty of Physics, Germany}

\author[0000-0003-2215-8485]{Renyu Hu} \affiliation{Jet Propulsion Laboratory, California Institute of Technology, Pasadena, CA 91109, USA} \affiliation{Division of Geological and Planetary Sciences, California Institute of Technology, Pasadena, CA 91125, USA}

\author[0000-0002-9428-8732]{Luigi Mancini} \affiliation{Department of Physics, University of Rome ``Tor Vergata'', Rome, Italy} \affiliation{Max Planck Institute for Astronomy, Heidelberg, Germany} \affiliation{INAF -- Turin Astrophysical Observatory, Pino Torinese, Italy \\}

\author[0000-0002-0502-0428]{Karan Molaverdikhani} \affiliation{Universit\"ats-Sternwarte, Fakult\"at f\"ur Physik,   Ludwig-Maximilians-Universit\"at M\"unchen, Scheinerstr.~1, 81679 M\"unchen, Germany}\affiliation{Exzellenzcluster `Origins', Boltzmannstr.~2, 85748 Garching, Germany \\}

\author[0000-0002-4262-5661]{Giuseppe Morello} \affiliation{Department of Space, Earth and Environment, Chalmers University of Technology, SE-412 96 Gothenburg, Sweden;}\affiliation{Instituto de Astrof\'isica de Canarias, E-38200 La Laguna, Tenerife, Spain}

\author[0000-0002-6500-3574]{Nikolay K. Nikolov} \affiliation{Space Telescope Science Institute, 3700 San Martin Drive, Baltimore, MD 21218, USA}

\author[0000-0001-8236-5553]{Matthew C.\ Nixon} \affiliation{Department of Astronomy, University of Maryland, College Park, MD, USA}

\author[0000-0003-3786-3486]{Seth Redfield} \affiliation{Astronomy Department and Van Vleck Observatory, Wesleyan University, Middletown, CT 06459, USA}

\author[0000-0002-7352-7941]{Kevin B. Stevenson} \affiliation{Johns Hopkins APL, Laurel,  MD, USA}

\author[0000-0003-4328-3867]{Hannah R. Wakeford} \affiliation{University of Bristol, HH Wills Physics Laboratory, Tyndall Avenue, Bristol, UK}

\author[0000-0003-4157-832X]{Munazza K. Alam} \affiliation{Carnegie Earth \& Planets Laboratory, 5241 Broad Branch Road NW, Washington, DC 20015, USA}

\author[0000-0001-5578-1498]{Björn Benneke} \affiliation{Department of Physics and Institute for Research on Exoplanets, Universite de Montr\'eal, Montreal, QC, Canada}

\author[0000-0002-0769-9614]{Jasmina Blecic} \affiliation{Department of Physics, New York University Abu Dhabi, Abu Dhabi, UAE} \affiliation{Center for Astro, Particle and Planetary Physics (CAP3), New York University Abu Dhabi, Abu Dhabi, UAE}

\author[0000-0001-7866-8738]{Nicolas Crouzet}\affiliation{Leiden Observatory, Leiden University, P.O. Box 9513, 2300 RA Leiden, The Netherlands}

\author[0000-0002-6939-9211]{Tansu Daylan}\affiliation{Department of Astrophysical Sciences, Princeton University, 4 Ivy Lane, Princeton, NJ 08544}\affiliation{LSSTC Catalyst Fellow}

\author[0000-0001-9164-7966]{Julie Inglis}\affiliation{Division of Geological and Planetary Sciences, California Institute of Technology, Pasadena, CA, 91125}

\author[0000-0003-0514-1147]{Laura Kreidberg}\affiliation{Max Planck Institute for Astronomy, Heidelberg, Germany}

\author[0000-0002-8963-3810]{Dominique J.M. Petit dit de la Roche}\affiliation{D\'{e}partement d’Astronomie, Universit\'{e} de Gen\`{e}ve, Sauverny, Switzerland}

\author[0000-0001-7836-1787]{Jake D. Turner}\altaffiliation{NHFP Sagan Fellow}\affiliation{Department of Astronomy and Carl Sagan Institute, Cornell University, Ithaca, New York 14853, USA}



\begin{abstract}



Carbon monoxide was recently reported in the atmosphere of the hot Jupiter WASP-39b using the 
NIRSpec PRISM transit observation of this planet, collected as part of the JWST Transiting Exoplanet Community Early Release
Science (JTEC ERS) Program. This detection, however, could not be confidently confirmed in the initial analysis of the higher resolution 
observations with NIRSpec G395H disperser. Here we confirm the detection of CO in the atmosphere of WASP-39b using the NIRSpec G395H data and cross-correlation techniques. We do this by searching for the CO signal in the unbinned transmission spectrum of the planet between 4.6 and 5.0 $\mu$m, where the contribution of CO is expected to be higher than that of other anticipated molecules in the planet's atmosphere. Our search results in a detection of CO with a cross-correlation function (CCF) significance of $6.6 \sigma$ when using a template with only ${\rm ^{12}C^{16}O}$ lines. The CCF significance of the CO signal increases to $7.5 \sigma$ when including in the template lines from additional CO isotopologues, with the largest contribution being from ${\rm ^{13}C^{16}O}$. Our results highlight how cross-correlation techniques can be a powerful tool for unveiling the chemical composition of exoplanetary atmospheres from medium-resolution transmission spectra, including the detection of isotopologues.


\end{abstract}



\section{Introduction} \label{sec:intro}

 The detection of carbon monoxide ($\rm CO$) and water ($\rm H_2O$) in gas giant exoplanet atmospheres has long been regarded as key to  understanding how these objects form. From atmospheric $\rm CO$ and $\rm H_2O$  measurements it is possible to estimate the planets carbon-to-oxygen (C/O) ratios. 
 The observed C/O ratios can then help trace the location in the stellar protoplanetary disk where the planets formed, and how they accreted the solids and gas that compose their cores and their atmospheres \citep[see e.g.,][]{Oberg2011, Mordasini2016, Madhusudhan2017, Cridland2019}. Isotopologue ratios have also been suggested as good tracers of planet formation conditions and atmospheric evolution \cite[e.g.,][]{ClaytonNittler2004,MolliereSnellen2019}.
 
$\rm H_2O$ has been readily detectable in planetary atmospheres for years, given the Hubble Space Telescope ($\rm HST$) capacity to observe one of its main bands centered around 1.4 $\mu$m \citep[e.g.,][]{Deming2013, Crouzet2014, Wakeford2018, Benneke2019}. $\rm CO$, on the other hand, has been harder to detect because of the lack of suitable space-based instrumentation targeting the spectral ranges of interest. Nevertheless, a few detections have been possible from the ground for exoplanets around very bright stars, using high-resolution observations (R $\sim$ $\rm 30,000-100,000$) centered on the $\rm CO$ 2.3 $\mu$m band \citep[e.g.,][]{Snellen2010_CC,Rodler2012,Brogi2012, deKok2013, Brogi2014,vanSluijs2022}, and for young exoplanets via direct imaging \citep[e.g.,][]{Konopacky2013,Barman2015,Petit2018}. Recent studies have also reported ${\rm ^{12}C^{16}O/^{13}C^{16}O}$ ratios in the atmospheres of a young accreting super-jupiter \citep{Zhang2021A} and a brown dwarf \citep{Zhang2021B}, using high-resolution observations centered on the CO 2.3$\mu$m band.

With the successful launch of $\rm JWST$, it is now possible to observe the CO band at 2.3 $\mu$m and its fundamental band centered at 4.7 $\mu$m with several of $\rm JWST$'s onboard instruments. This, combined with $\rm JWST$'s capacity to observe fainter targets than previous missions could, opens the possibility to search for $\rm CO$ in a large number of exoplanets, enabling future composition population studies. 

Early $\rm JWST$ observations have already produced detections of $\rm CO$ in exoplanetary atmospheres. Specifically,  \cite{Miles2022} detected the 2.3 $\mu$m and 4.7 $\mu$m CO bands in the young, directly imaged gas giant planet VHS 1256-1257 b using the NIRSpec IFU setup, and \cite{Rustamkulov22} reported a detection of CO at 4.7 $\mu$m in the transiting gas giant exoplanet WASP-39b using NIRSpec PRISM \footnote{Low-resolution (R $\sim$ $20-300$) double-pass prism disperser.} (R $\sim$ $20-300$). This last CO detection could not be confirmed by the initial analysis of the observations of WASP-39b with NIRSpec G395H \footnote{High-resolution (R $\sim$ $2700$) grating disperser.} \citep{Alderson2022}, although a more recent analysis by \cite{Grant2023} has detected CO in these data using a new technique that searches for the CO signal in pre-defined sets of wavelength sub-bands. 

WASP-39b \citep{Faedi11} is a 0.28~$\rm M_{Jup}$, 1.27~$\rm R_{Jup}$
exoplanet with an equilibrium temperature of about 1200\,K, orbiting a G7\,V star. Four transits of WASP-39b were observed as part of the JWST Director's Discretionary Time Transiting Exoplanet Community Early Release Science Program \citep[ERS-1366 PI: N. Batalha;][]{Stevenson2016,Bean2018} with the NIRSpec PRISM and G395H, NIRCam WFSS (Wide Field Slitless Spectroscopy), and NIRISS SOSS (Single Object Slitless Spectroscopy) instrument modes, which combined detected a number of chemical species, including  Na ($19\sigma$), H$\rm_{2}$O ($33\sigma$), CO$\rm_{2}$ ($28\sigma$), CO ($7\sigma$) \citep{Rustamkulov22,Alderson2022, Ahrer2022, Feinstein2022}
and the unexpected photochemical byproduct SO$\rm_{2}$ ($4.8\sigma$) \citep{Tsai2022, Alderson2022}. The detection of CO and CO$\rm_{2}$, combined with the absence of CH$\rm_{4}$ in the transmission spectrum of WASP-39b, reveals that the planet has a super-solar metallicity.

Here we set out to confirm the detections of $\rm CO$ in \cite{Rustamkulov22} and \cite{Grant2023} and search for CO isotopologues using the JTEC ERS Program NIRSpec G395H dataset and cross-correlation techniques.  Cross-correlation techniques have been ubiquitously used in astronomy for decades \citep[e.g.,][]{Tonry1979}, but they have only been used for exoplanet atmospheres studies through transmission spectroscopy at high ($R \sim 100,000$) resolutions  \cite[e.g.,][]{Snellen2010_CC} or direct imaging at medium ($R \sim 1,000$) resolutions \cite[e.g.,][]{Konopacky2013}. The superb stability and data quality of JWST observations allows us to apply this technique to medium resolution transmission spectra.

\begin{figure*}[th!]
\includegraphics[width=\textwidth]{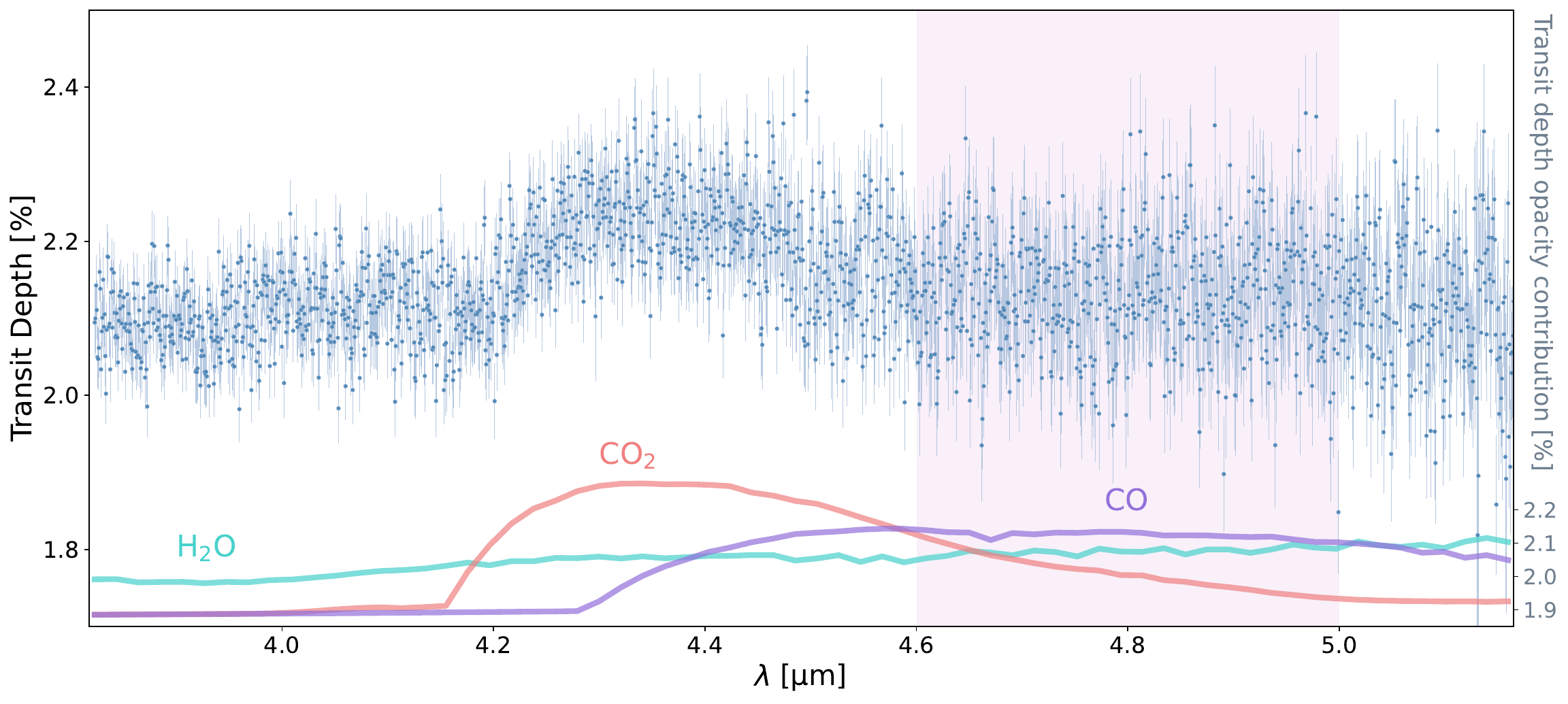}
\caption{Transmission spectrum of WASP-39b obtained from the JTEC ERS Program NIRSpec G395H NRS2 observation using the \texttt{Tiberius} pipeline. In the lower part of the plot and in the right y-axis, we show the relative opacity contributions of $\rm H_{2}O$, $\rm CO$ and $\rm CO_{2}$, which are the main features expected in that wavelength range. The vertical, purple-shaded area highlights the part of the spectrum where the contribution of $\rm CO$ is expected to be predominant over $\rm H_{2}O$ and $\rm CO_{2}$. That is the portion of the spectrum used in the CCF analysis.}
\label{fig:All_data_subfig}
\end{figure*}

\section{Observations and data reduction} \label{sec:data}


JWST observed a full transit of WASP-39b with the NIRSpec G395H mode \citep{Jakobsen22,Birkmann22} as part of the JTEC ERS Program.
The target was monitored for 8 hours and 36 min between 30 July 2022 21:45 and 31 July 2022 06:21 UTC, using the Bright Object Time Series mode (BOTS), with the 1.6$\arcsec$ $\times$ 1.6$\arcsec$ slit, SUB2048 subarray, and the NRSRAPID readout pattern. The observation resulted in a total of 465 integrations, with 70 groups per integration, with the spectra of the target dispersed across the NRS1 and NRS2 detectors. A particular feature of this dataset is a \textit{tilt event} that produced a mid-transit jump in flux \citep[see][for details]{Alderson2022}.

We reduced the NRS1 and NRS2  spectra separately using the \texttt{Tiberius} pipeline  \citep{2017MNRAS.468.3907K,Kirk18,Kirk19,Kirk21}, but here we focus on the NRS2 detector, which covers the CO fundamental band. \texttt{Tiberius} works from the group-level 2D \textit{rateints} images produced by the STScI Science Calibration Pipeline \citep[][ v1.6.2]{Rigby22}, from which we remove the \textit{1/f noise} contributions \citep{Jakobsen22,Birkmann22} per column using the median of all off-target pixels in each column.  Images were corrected for bad and saturated pixels manually, using custom-made masks that flagged all pixels deviating more than 5$\sigma$ from running medians along pixel rows and replacing their count values by interpolating the counts from nearby pixels. We traced the spectra fitting Gaussian distributions at each column, and then extracted each spectrum using 4-pixel wide apertures centered on the trace. Residual background flux not captured by the \textit{1/f} correction was removed by fitting linear polynomials along each column of the detector after masking a 3-pixel-wide region centered on the trace. To remove residual cosmic rays in each stellar spectrum, we compared each spectrum to the median of all the spectra, iteratively removing 8$\sigma$ or larger outliers and replacing them by linear interpolation with neighboring pixels. Finally, we corrected for any potential wavelength shift in each spectrum by cross-correlating them with the first spectrum and putting all the spectra in a common wavelength grid by linearly resampling them. We used wavelength maps from the JWST Calibration Pipeline to produce the wavelength solution, as in \cite{Alderson2022}.


We produced a white light curve (WLC) from the extracted spectra by summing the flux per spectrum over the full wavelength range of the detector (3.823--5.177\,$\mu$m), and fitted it with a transit + Gaussian Processes (GP) systematics model using \texttt{batman} \citep{kreidberg2015}, \texttt{george} \citep{ambikasaran2015}, and a Monte Carlo Markov Chain (MCMC) implemented using \texttt{emcee} \citep{foreman-mackey2013}.  The model fitted for the planet-to-star radii ratio (${\rm R_p/R_s}$), orbital inclination (${\rm i}$), epoch of transit (${\rm T_0}$), semimajor axis in units of the stellar radius (${\rm a/R_s}$), the ${\rm u_1}$ quadratic law
limb-darkening coefficient, and an exponential squared GP kernel with the x- and y-pixel shifts, the FWHM of the spectra, and sky background as parameters. We used a Gaussian prior centered on 0.015 to fit for ${\rm a/R_s}$, and uniform priors for all other fitted parameters. For this fit, we kept the orbital period fixed to the value reported by \cite{Mancini18} and the eccentricity fixed to zero. We fixed the ${\rm u_2}$ limb darkening coefficient to the value computed by  ExoTIC-LD for the quadratic limb darkening law. 

For the spectroscopic light curves (SLC), we generated 2044 light curves at 1-pixel resolution within the NRS2 wavelenth range and fitted them using a similar procedure to the one described above, but holding
 ${\rm i}$, ${\rm T_0}$  and ${\rm a/R_{s}}$ fixed to the values obtained from the 
WLC fit, and applied a systematics correction from the WLC fit to aid in fitting the mirror tilt event.  We then fitted each spectroscopic light curve using polynomials to describe residual systematics, not GPs. We chose a combination of four linear polynomials describing the change in x-position and y-position of the star on the detector, FWHM of the stellar trace and the measured background flux.  

Table~\ref{tab:adopted_parameters} summarizes the adopted system parameters, the priors for the parameters fitted in the white and spectroscopic light curves, and the posteriors for the WLC fit. Figure~\ref{fig:All_data_subfig} shows the obtained transmission spectrum of WASP-39b between 3.823--5.177\,$\mu$m. 
The data in the figure are provided in Table~\ref{tab:transmission_spectrum}.

We note that in our analysis we use the spectrum extracted per pixel column, not per resolution element as  defined by the Nyquist sampling limit\footnote{JWST User Documentation (JDox). Baltimore, MD: Space Telescope Science Institute; \url{https://jwst-docs.stsci.edu}}, since both samplings produce consistent results, but the \textit{per pixel column} sampling is more sensitive to the signal of CO and its isotopologues as demonstrated in Appendix~\ref{app: sampling}.



\begin{table}[!ht]
  \caption{Wavelength range per point, transit depth, TD, and transit depth uncertainty, $\Delta$TD, of the WASP-39b transmission spectrum derived using \texttt{Tiberius}. 
  The full table is available electronically in the online version of this article.\label{tab:transmission_spectrum}}
  \centering
  \begin{tabular}{cccc}
  \toprule \toprule
    $\lambda_{1}$ ($\mu$m) & $\lambda_{2}$ ($\mu$m) &  TD (ppm) &  $\Delta$TD (ppm) \\
    \midrule
3.822427 & 3.823099 & 0.02095 & 0.00044 \\
3.823099 & 3.823770 & 0.02100 & 0.00045 \\
3.823770 & 3.824442 & 0.02143 & 0.00040 \\
... & ... & ... & ...  \\
    \bottomrule \bottomrule
           
  \end{tabular}
\end{table}

\section{CO Template Generation}\label{sec:COtemplate}

The fundamental band of CO covers air wavelengths between 4.28 and 6.50 $\mu$m. The portion of the band in the rest frame of WASP-39b covered by the NIRSpec NRS2 detector is highlighted by the purple line in Fig.~\ref{fig:All_data_subfig}. For our search, however, we limit the spectral range to 4.6 -- 5.0 $\mu$m 
to minimize contamination from ${\rm CO_2}$ below 4.6 $\mu$m and ${\rm H_2O}$ above 5.0 $\mu$m, where the signals from those molecules dominate.

WASP-39 has a barycentric systemic velocity of $\rm-58.4421$ ${\rm km~s^{-1}}$ \citep{Mancini18}, and the JWST barycentric velocity at the time of the observations was -28.888 ${\rm km~s^{-1}}$, as recorded in the image headers under the keyword \textsc{velosys}. Therefore, any atmospheric ${\rm CO}$ signal from WASP-39b is expected to appear shifted by about -87.330 ${\rm km~s^{-1}}$. 



We generated pure CO atmospheric transmission spectrum 
templates of WASP-39b with \texttt{petitRADTRANS} \citep{Petitradtrans}, using the \textit{lbl} (line-by-line) radiative transfer mode with a resolving power of $\lambda/\Delta \lambda=10^{6}$, and adopting both the main $\rm ^{12}C^{16}O$ isotopologue and the \textit{all CO isotopes}\footnote{ i.e., $^{12}$C$^{16}$O,$^{13}$C$^{16}$O,$^{12}$C$^{18}$O,$^{12}$C$^{17}$O,$^{13}$C$^{18}$O and $^{13}$C$^{17}$O following abundances of $98.65~\%$, $1.11~\%$, $0.20~\%$, $3.7\times10^{-2}~\%$, $2.2\times10^{-3}~\%$ and $4.1\times10^{-4}~\%$, respectively.} high-resolution line lists \textit{"CO\_main\_iso"}  
 and \textit{"CO\_all\_iso"} from HITEMP \citep{HITEMP}. 
 For the planet, we adopted a radius of 1.27 $\rm R_{Jup}$, a surface gravity of 407.38 cm~$\rm s^{-2}$ following \cite{Faedi11}, an isothermal temperature profile of 1166 K \citep{Mancini18}, and atmospheric bottom and top pressures of  $10^{2}$ and $10^{-12}$ bars, respectively. 

\begin{figure*}[ht!]
\epsscale{1.2}
\gridline{\fig{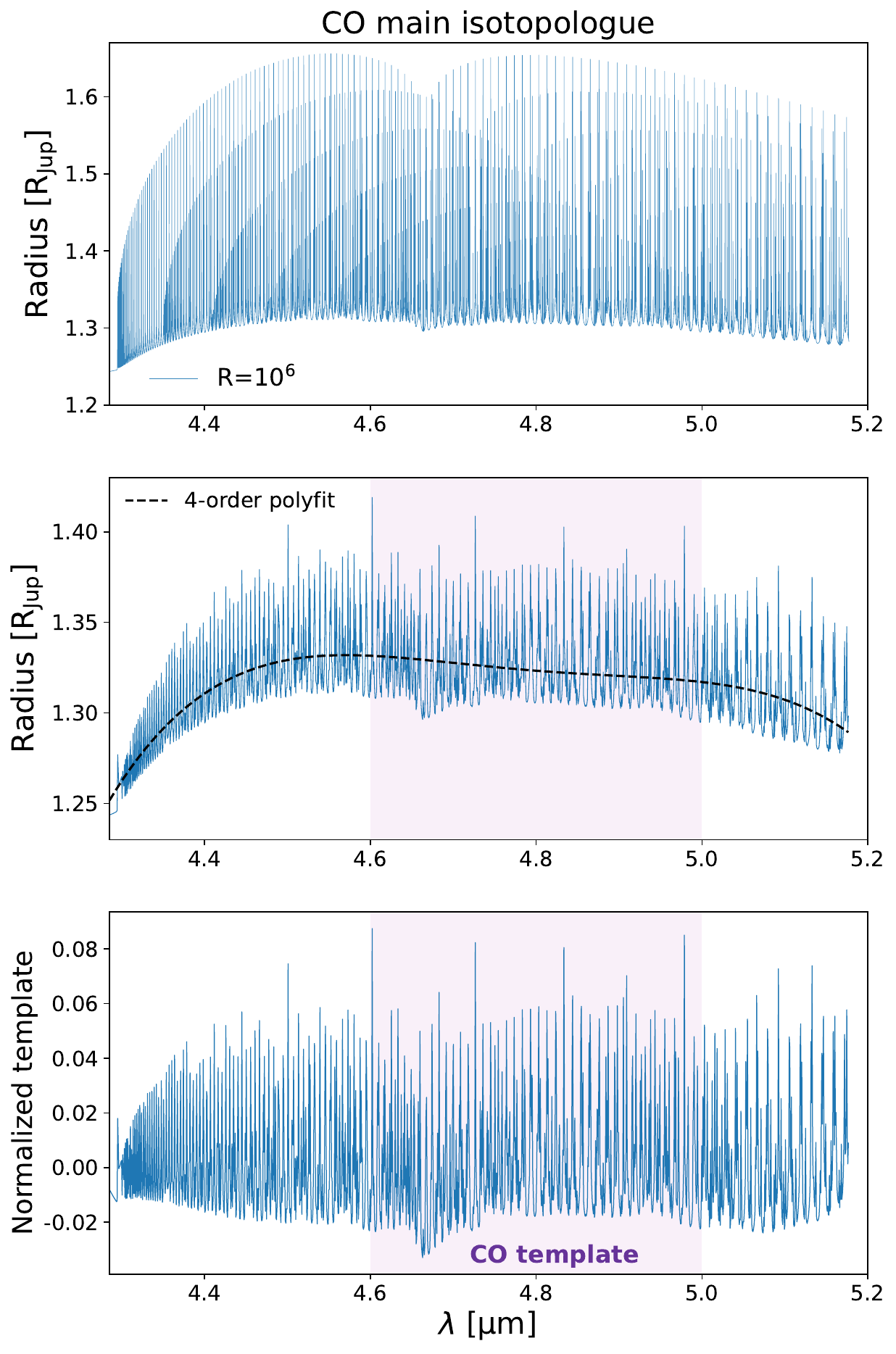}{\columnwidth}{(a)}\fig{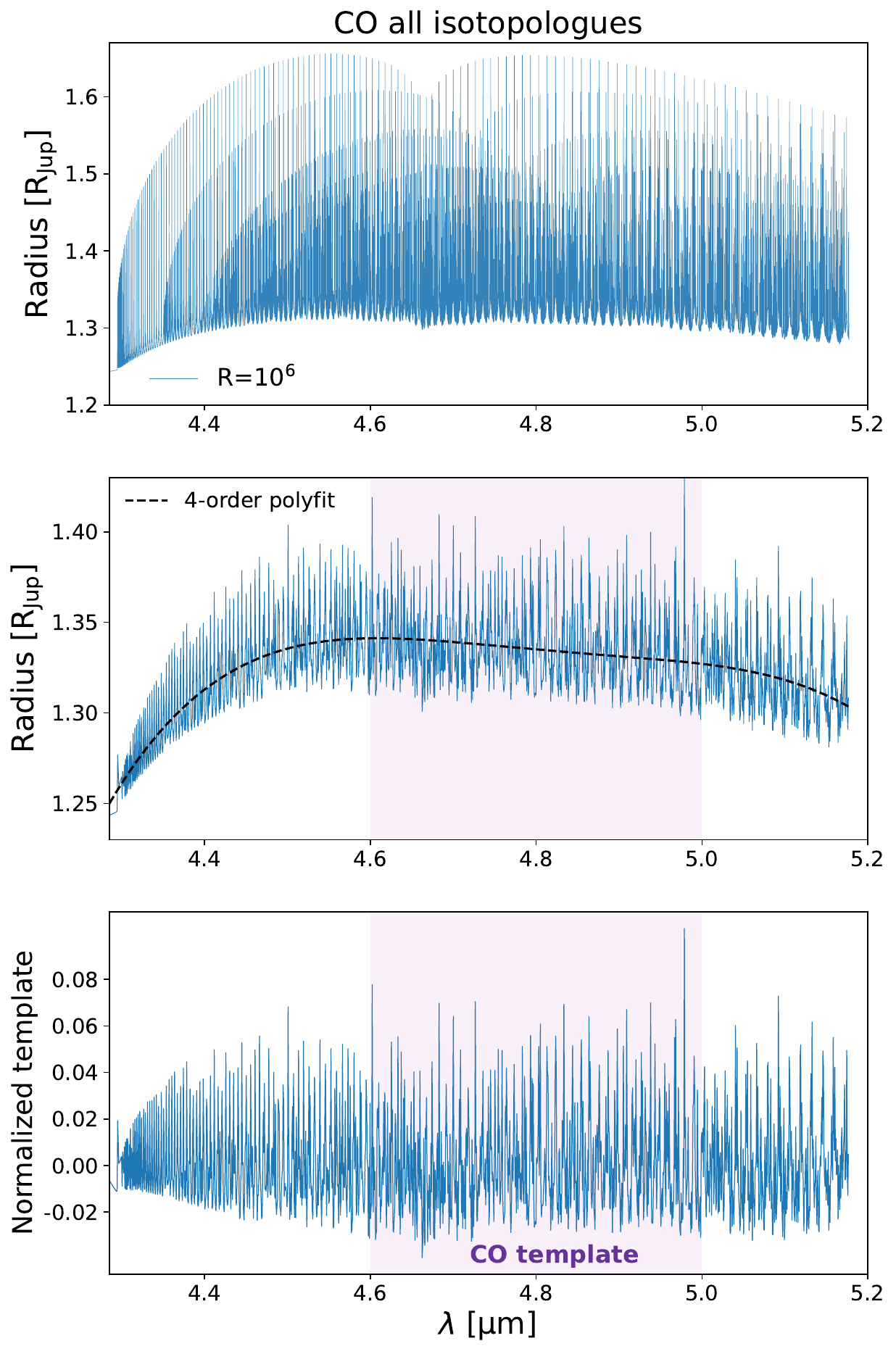}{\columnwidth}{(b)}}
\caption{Atmospheric models for the main CO isotopologue (a) and all CO isotopologues (b). Top: High-resolution ($10^{6}$) \texttt{petitRADTRANS}-computed transmission spectrum model of an atmosphere composed solely of CO. Middle: CO spectrum model degraded to match the resolution of the observed transmission spectrum. The region used for the cross-correlation search is highlighted in purple, with resolving powers of ${\rm R = 3470}$ and ${\rm3803}$ between 4.6 and 5.0 $\rm \mu$m, respectively. The black dashed line shows the 4-order polynomial fit that is used for the normalization. Bottom: CO template used in the cross-correlation analysis, calculated by subtracting the 4-order polynomial fit to the ${\rm R = 3470-3803}$ CO model.
\label{fig:Atmospheric_CO_models}}
\end{figure*}

The top panels of Figure~\ref{fig:Atmospheric_CO_models} show the resultant R=$\rm 10^{6}$ ${\rm CO}$ model spectra for the main CO isotopologue (left) and for all CO isotopologues (right). To match the resolution of the observations, we resampled the template ${\rm CO}$ spectrum produced by \texttt{petitRADTRANS} to the sampling of the observations using \texttt{spectres}\footnote{\url{https://spectres.readthedocs.io/}} \citep{spectres}, which produced the spectra 
shown in the middle panel of Fig.~\ref{fig:Atmospheric_CO_models}. The bottom panel of Fig.~\ref{fig:Atmospheric_CO_models} shows the resampled templates, normalized using a 4th-order polynomial fit. The portion  of these templates between 4.6 and 5.0 $\rm \mu$m, highlighted in purple, is the one used in the cross-correlation search for CO.

As a final step  before starting the cross-correlation search, we normalized the observed transmission spectrum of WASP-39b between 4.6 and 5.0\,$\mu$m by subtracting from the spectrum its mean value in that wavelength range. 

\section{Cross-Correlation Search Results}\label{sec:CCSearch}










We carried out cross-correlations between the normalized observed spectrum and the normalized CO \textit{main-isotope} and CO \textit{all-isotopes} templates using the \textit{crosscorr} function in \texttt{PyAstronomy} \citep{pyastronomy}\footnote{\url{https://github.com/sczesla/PyAstronomy}}, shifting the templates with respect to the observed spectrum over a range of systemic velocities between $\pm$ 1000 ${\rm km~s^{-1}}$, in 6 ${\rm km~s^{-1}}$ steps, which is about 15 times smaller than the pixel-to-pixel radial-velocity sampling of the spectrum. We also tested the cross-correlation using a smaller, 1 ${\rm km~s^{-1}}$ step, obtaining analogous results. 
We calculated uncertainties for each correlation value (CV) by propagating the uncertainties in the observed transmission spectrum of WASP-39b as
\begin{equation}
    \rm
    \Delta CV(RV) = \sum_{\lambda} \Delta S(\lambda) T(RV,\lambda) ~,
\end{equation}

\noindent where $\rm \Delta CV(RV)$ is the uncertainty of each point of the cross-correlation function (CCF), $\rm \Delta S(\lambda)$ is the uncertainty of each point of the transmission spectrum, and $\rm T(RV,\lambda)$ is the template used in the cross-correlations at each RV shift.

\begin{figure*}[ht!]
\gridline{\fig{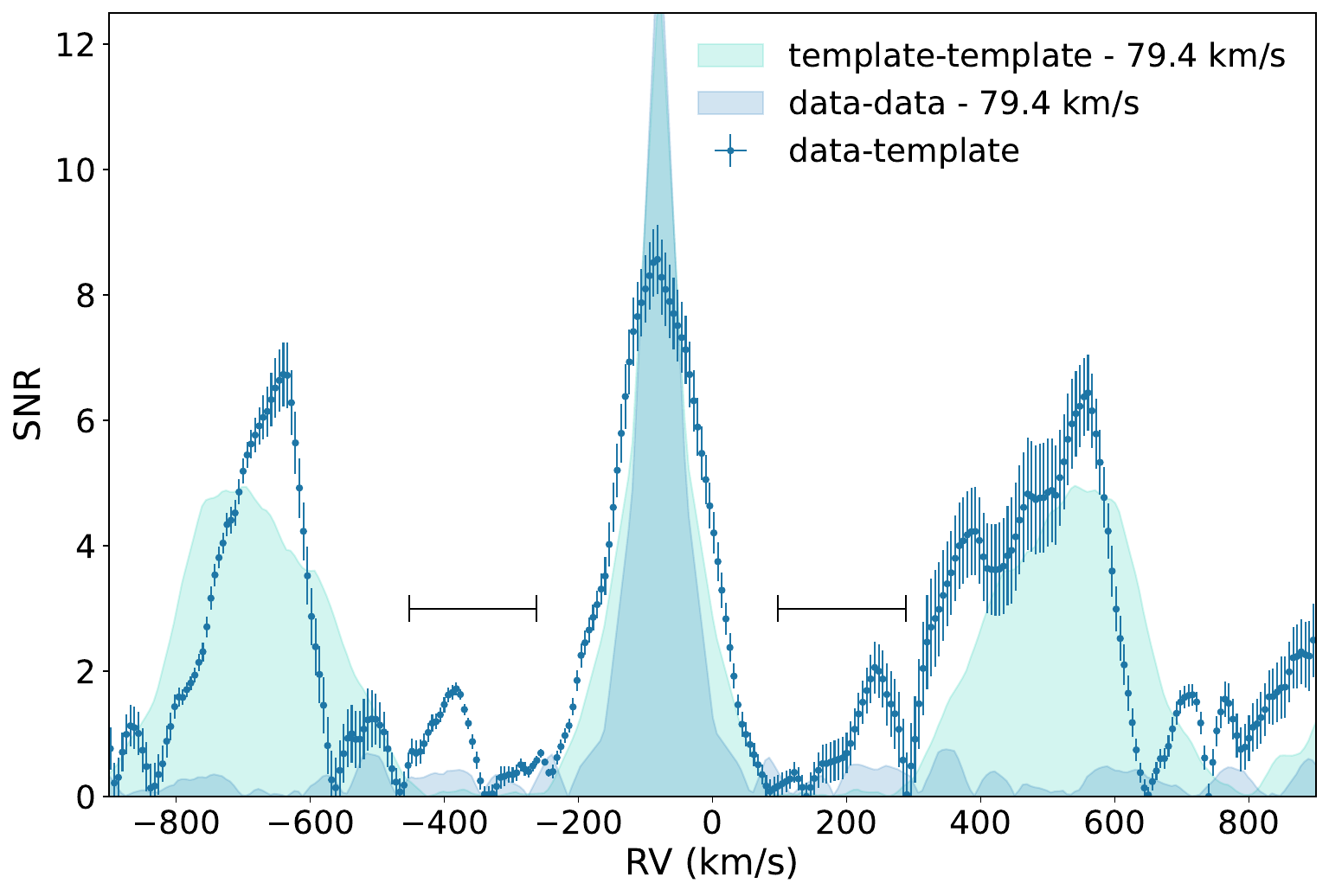}{\columnwidth}{(a) CO main isotopologue}\fig{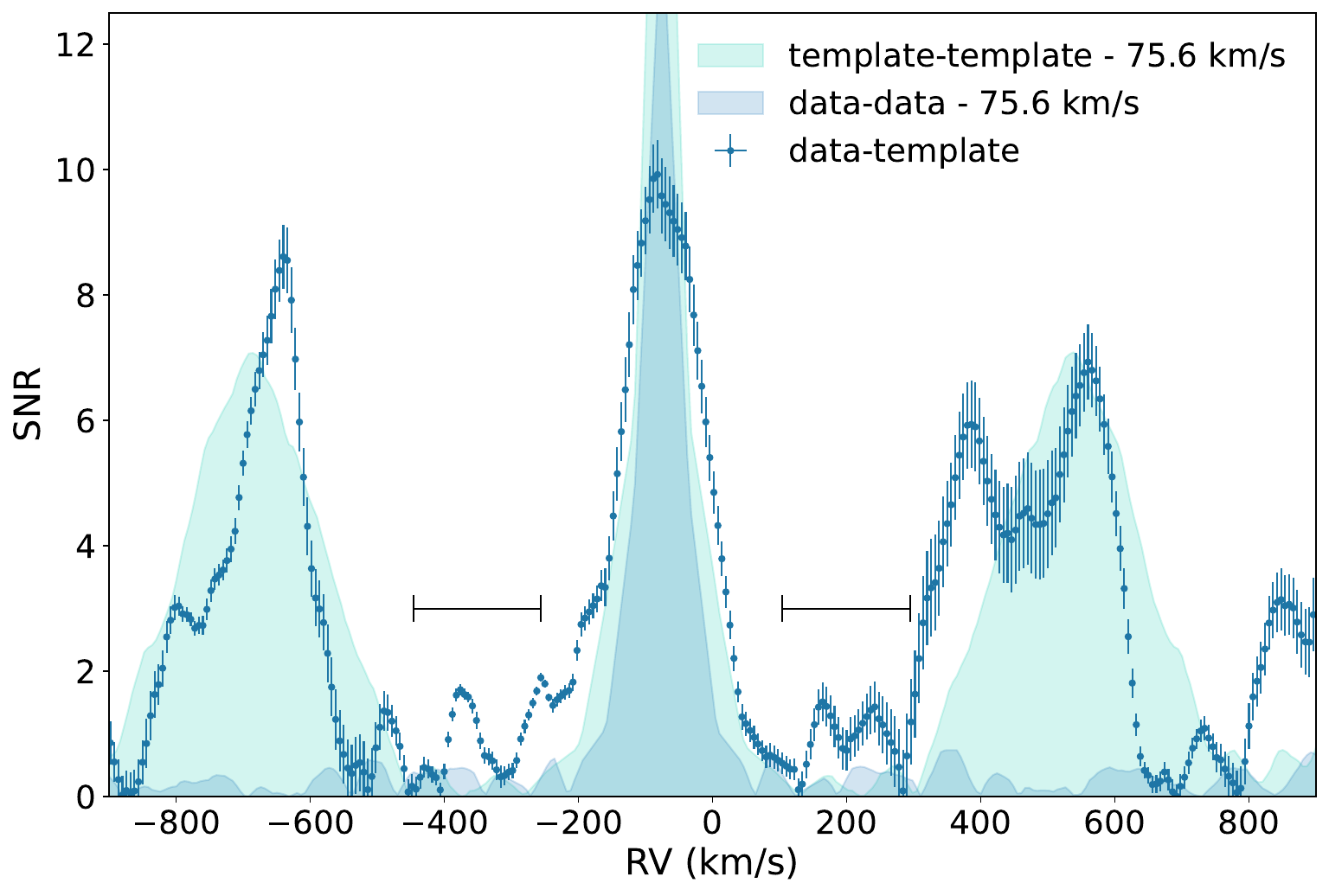}{\columnwidth}{(c) CO all isotopologues}}
\gridline{\fig{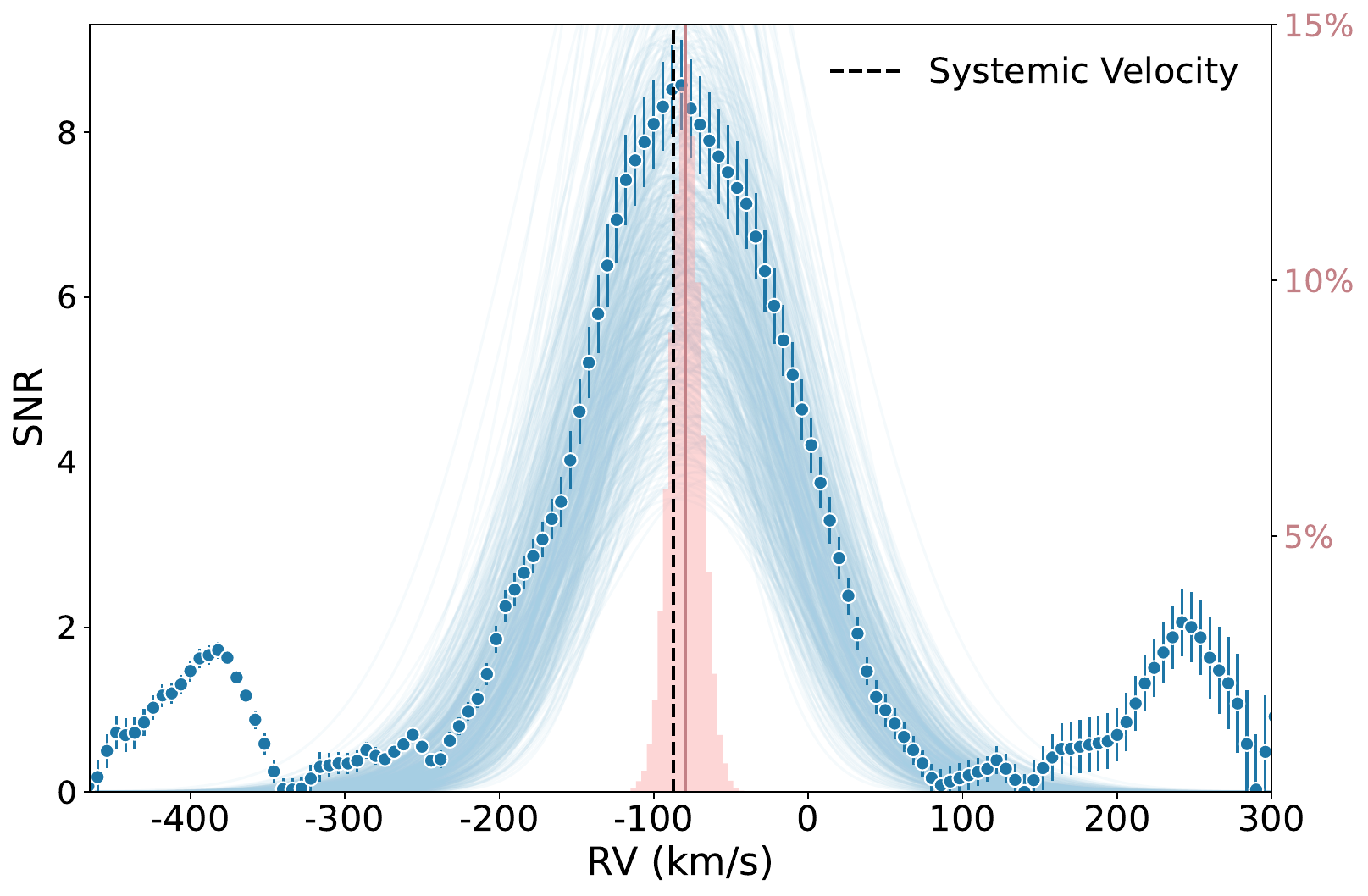}{\columnwidth}{(b) CO main isotopologue}\fig{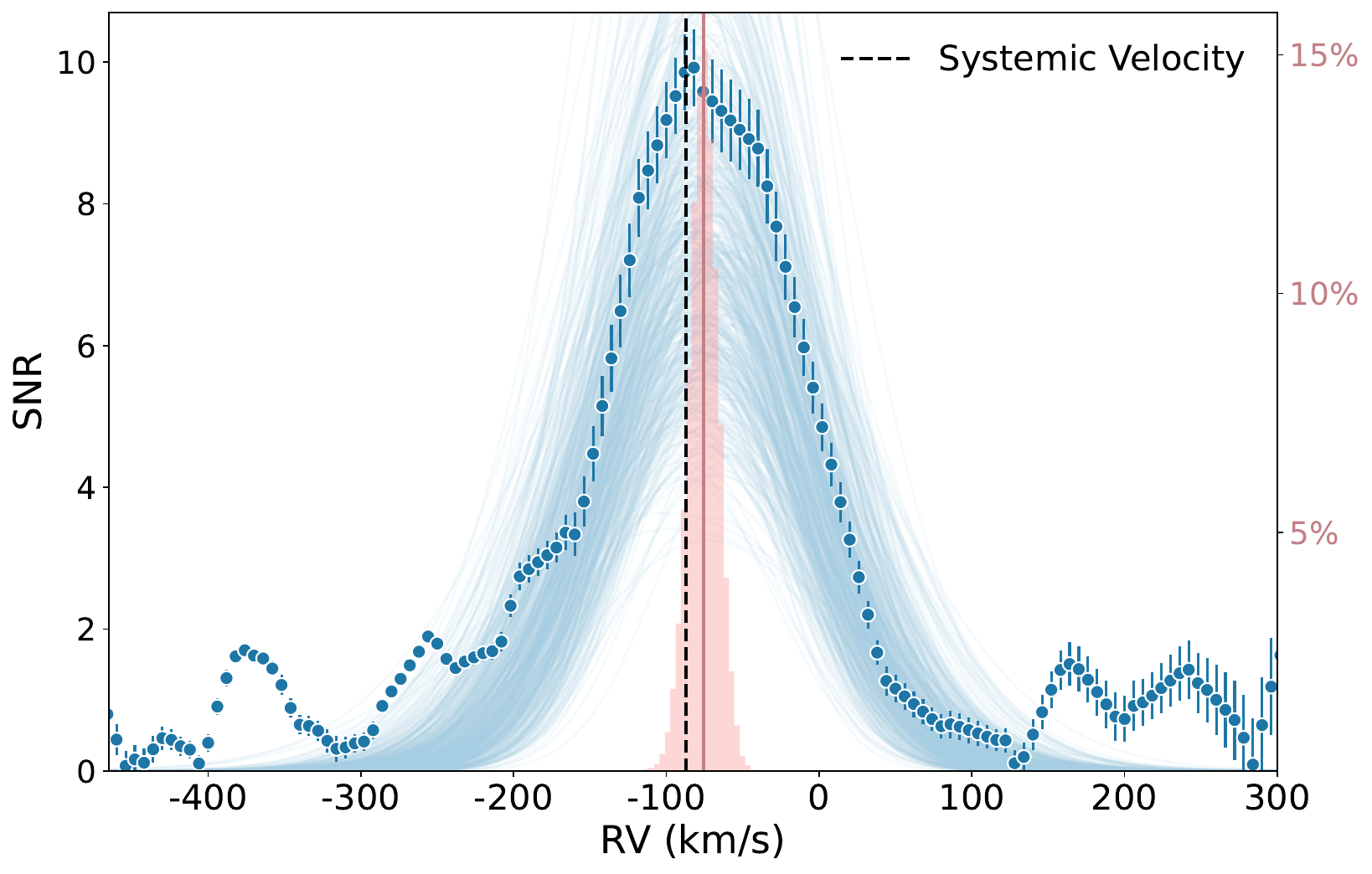}{\columnwidth}{(d) CO all isotopologues}}
\caption{Cross-correlation function (CCF) of the WASP39b transmission spectrum with the CO main isotopologue (left panels) and the CO all isotopologues templates (right panels) in the 4.6--5.0\,$\rm \mu$m range using a radial velocity step of 6~km/s (blue dotted line). The black horizontal bars indicate the continuum regions used to estimate the noise of the CCF, and calculate the significance of the central peak. In panels (a) and (c) we compare the CCFs resulting from the self cross-correlation of the template (light blue area) and the spectrum (dark blue area), which were scaled arbitrarily. Panels (b) and (d) show the central peak of the CCF compared to the histogram of the measurements of the CCF peak radial velocity through random sampling. The right y-axis shows the percentage of samples. The red-continuum line shows the mean value of the RV measurements distribution. The black-dashed line shows the systemic velocity relative to the JWST at the time of the observation. The light-blue lines show a set of Gaussian fits among the sampled CCFs. 
\label{fig:sigmas_mainiso}}
\end{figure*}

To put the resulting CVs into more tangible units, 
we converted the CVs to signal-to-noise ratio (SNR) values as follows: We defined out-of-peak \textit{continuum} regions where the CVs 
are expected to be dominated by noise. These regions are between [$-452$, $-262$]~${\rm km~s^{-1}}$ and [$+98$, $+288$]~${\rm km~s^{-1}}$, defined as 190~${\rm km~s^{-1}}$ wide regions on each side of the central peak of the CCF, starting at velocities $\pm$ 3$\sigma$ away from that central peak, assuming that the distribution of velocities in the peak is Gaussian, see Fig.~\ref{fig:sigmas_mainiso}. The conversion of the CVs to SNR values is then calculated as

\begin{equation}
    \rm
    SNR(per~point) = \frac{|CV(per ~point) - \mu_{cont}|}{\sigma_{cont}} 
    ~,
\end{equation}

\noindent where  $\mu_{cont}$ is the mean of the CVs in the continuum regions defined above, and $\sigma_{cont}$ is the standard deviation of the CVs in those regions. 

The resultant CCFs in SNR units are shown in Figure \ref{fig:sigmas_mainiso}, where the continuum regions are indicated by the horizontal black lines in the top panels.
The CCFs for the CO \textit{main-isotopologue} and the CO \textit{all-isotopologues} templates both show SNR $>$ 8 peaks centered near $-80 ~{\rm km~s^{-1}}$. This velocity shift is consistent with the barycentric velocity of the system at the time of the observations as discussed in Sect.~\ref{sec:COtemplate}.




To measure the significance and the actual central velocity of the main  peaks in each CCF we fit a Gaussian to the SNR-converted CCF using the least squares routine \textit{scipy.optimize.curvefit}, and fit for the SNR value of the central peak, ${\rm S_{peak}}$, its central velocity, ${\rm RV_{peak}}$, and the FWHM of the distribution of velocities, ${\rm \sigma_{RVs}}$.

\begin{figure*}[ht!]
\gridline{\fig{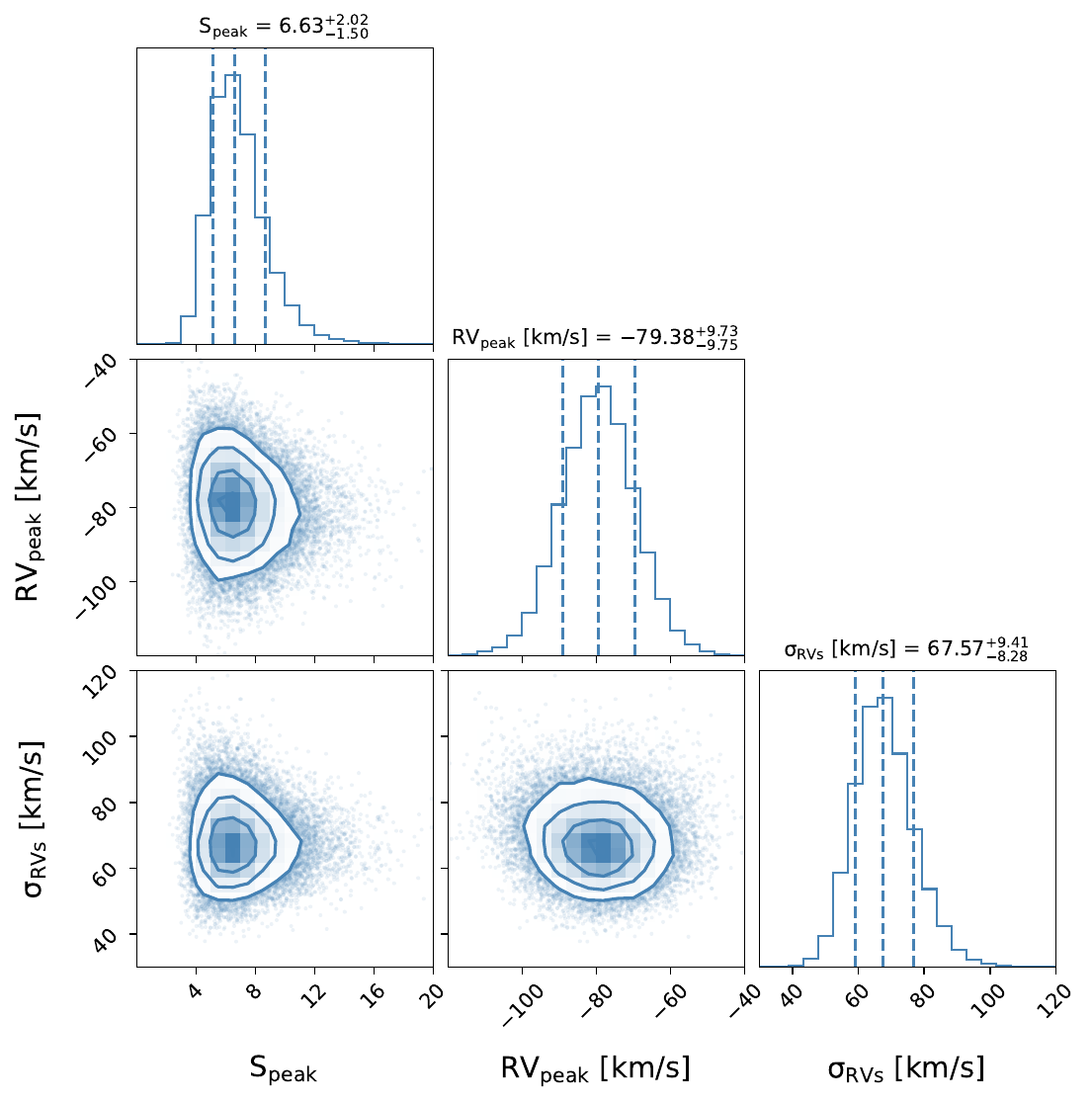}{\columnwidth}{(a) CO main isotopologue}\fig{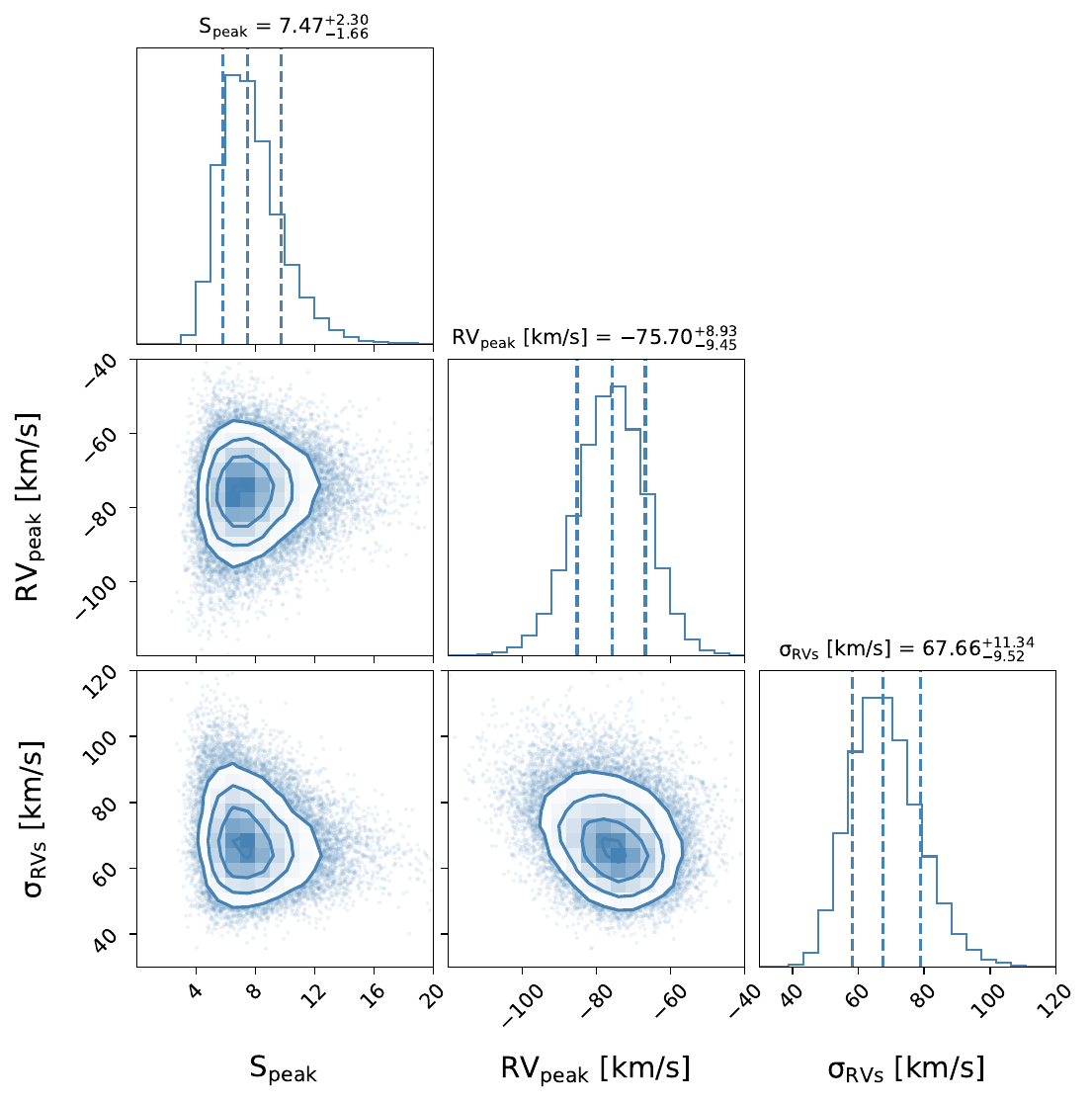}{\columnwidth}{(b) CO all isotopologues}}
\caption{Distribution of the best-fit Gaussian parameters (significance of the central peak, ${\rm S_{peak}}$, the central peak velocity, ${\rm RV_{peak}}$, and the FWHM of the distribution of velocities, ${\rm \sigma_{RVs}}$) retrieved from the sampled transmission spectra that were randomly generated following a Gaussian distribution centered in the observed transmission spectrum and considering the errors as the Gaussian standard deviation in each point. We performed 30000 iterations. Panels (a) and (b) show the results using the main CO isotopologue and all CO isotopologues templates, respectively.
\label{fig:corner}}
\end{figure*}

As a last step, we estimate uncertainties on the values of ${\rm S_{peak}}$ and ${\rm RV_{peak}}$,  calculated as described above, by generating 30,000 Monte Carlo random samples of the observed transmission spectrum of WASP-39b in Figure \ref{fig:All_data_subfig}, where each point is simulated using a uniform probability distribution of values within the range of errors observed in the spectrum. We then calculate cross-correlations for each simulated spectrum, with each of the two CO templates and measure the significance of the peak and central velocity of each peak as described above. The bottom panels of Figure \ref{fig:sigmas_mainiso} show the central CCF peaks for each CO template with a sample of the CCFs obtained from individual MCMC realizations in light blue, and the resultant histograms of the central radial velocities of the peak in red. The solid blue lines in the figures correspond to the CCFs obtained from the observed spectrum, and vertical dashed lines correspond to the velocity of the WASP-39 system at the time of the observations. The retrieved distributions of the three parameters, ${\rm S_{peak}}$, ${\rm RV_{peak}}$, and ${\rm \sigma_{RV}}$ for each CO template are shown in Figure \ref{fig:corner}.

Our analysis reveals CCF peaks centered at ${\rm -79.4 \pm 9.8~km~s^{-1}}$ with a significance of ${\rm 6.6~_{-1.50}^{+2.02}~\sigma}$ when using the CO \textit{main-isotopologue} template, and at ${\rm -75.7~ ^{+9.2}_{-9.4}~km ~s^{-1}}$ with a significance of ${\rm 7.5~_{-1.66}^{+2.30}~ \sigma}$ when using the CO \textit{all-isotopologues} template. Both velocities are consistent within about $1\sigma$ with the computed velocity of the system at the time of the observations of ${\rm -87.33~km ~s^{-1}}$. In Appendix~\ref{app: 13C16O}, we show an additional analysis of the observed transmission spectrum using a template composed of the two most abundant CO isotopologues: $\rm^{12}C^{16}O$ and $\rm^{13}C^{16}O$. The CCF results are similar to the ones obtained using the CO \textit{all-isotopologues} template. The significance of the detection, being $\rm 7.33^{+2.26}_{-1.04}$ sigma, increases compared to the CO \textit{main-isotopologue} result but is slightly less than the \textit{all-isotopologues} significance, indicating that the \textit{all-isotopologues} template produces a better match.


\subsection{Secondary Cross-Correlation Function Peaks}\label{sec:secondpeaks}

In addition to the central CCF peaks in agreement with the systemic velocity of WASP-39, the top panels in Figure~\ref{fig:sigmas_mainiso} show secondary CCF peaks at around -600 ${\rm km~ s^{-1}}$ and +600 ${\rm km~s^{-1}}$, which also appear significant.  To elucidate the origin of those peaks we cross-correlated the observed transmission spectrum of WASP-39b with itself, and each CO template with itself at the same resolution as the observations, over the same range of velocities, ${\rm \pm~1000~km~ s^{-1}}$, and with the same ${\rm 6~km~ s^{-1}}$ steps. The top panels in Figure \ref{fig:sigmas_mainiso} show the result of the template-to-template CCFs in cyan, and of the data-to-data CCF in light blue. The template-to-template and data-to-data CCFs have been shifted by the ${\rm RV_{peak}}$ values measured in the previous section so we can compare them directly to the data-to-template CCFs. 

The template-to-template CCFs show the expected peak at ${\rm 0~km~ s^{-1}}$, and two symmetric peaks at a mean distance of ${\rm \pm~624~km~ s^{-1}}$ from the central peak, similar to the ones observed in the data-to-template CCFs. These peaks are expected in the template-to-template CCFs of CO because of the periodic patterns in the distribution of lines in the spectrum of this molecule (see Figure \ref{fig:Atmospheric_CO_models}). The data-to-data CCF shows the expected peak at ${\rm 0~km~ s^{-1}}$, but no other significant peaks.  The presence of the same secondary peaks in the data-to-template and template-to-template CCFs reinforces the conclusion that the signal detected in the transmission spectrum of WASP-39b comes from CO.

\section{Conclusions and Discussion}\label{sec:conclusions}

We report the detection of CO in the atmosphere of WASP-39b via a cross-correlation analysis of the JTEC ERS Program NIRSpec G395H dataset. Our analysis produces a 6.6$\sigma$ CCF detection of the fundamental band of CO centered around 4.8 $\mu$m when using a $^{12}$C$^{16}$O 
template and a
 7.5$\sigma$ detection when using a template with all CO isotopologues included in \texttt{petitRADTRANS}. 

Our reported detection significance levels are relative to custom-designed baseline regions of the CCF, and as such cannot be readily used to estimate CO abundances. In addition, some stellar CO absortion is expected in WASP-39b's host star, as pointed out by \cite{Rustamkulov22}, and this may dilute the strength of the planetary CO signal, yielding to underestimated detection significances, an effect previously reported by \cite{Deming2017}.


Our result confirms the detections of CO in the atmosphere of WASP-39b reported by \cite{Rustamkulov22} and \cite{Grant2023}. 
This detection, combined with the detection of ${\rm CO_2}$ \citep{Ahrer22CO2}, and the absence of ${\rm CH_4}$ \citep{Alderson2022}, reinforces the conclusion of super-solar atmospheric metallicity for this planet \citep{Rustamkulov22,Alderson2022, Ahrer2022, Feinstein2022, Tsai2022}. 

In addition, we show that standard CCF analysis techniques can be used to identify the presence of molecular features in exoplanet atmospheres using transmission spectra with a resolving power of just a few thousand, as is the case of JWST observations in \textit{high}-resolution modes, provided that the spectra are of high-enough quality. CO has the advantage that, at the native resolution of JWST, it still presents a band shape with resolved lines. Still, this technique can be used to search for further molecules in the atmosphere of WASP-39b and also to identify atmospheric molecular species in other planets observed with JWST using these modes. 

We highlight the presence of secondary peaks in the CCFs described in section \ref{sec:secondpeaks}, which we also attribute to CO (see Appendix~\ref{app: sampling}). Even though we did not exploit the idea in this \textit{letter}, it is worth exploring if secondary peaks in CCFs can be used to increase the confidence of future CO detections, or other molecules, or to place stronger non-detection constraints. 

Finally, we highlight how the significance of our CO detection increases from 6.6$\sigma$ to 7.5$\sigma$ when adding a number of isotopologues to the $^{12}$C$^{16}$O template. 
  To evaluate the influence of individual isotopologues, we performed an additional cross-correlation analysis using a $\rm^{12}C^{16}O$ ($\rm 98.7\%$) + $\rm^{13}C^{16}O$ ($\rm 1.3\%$) template.  
  In this case the significance of the CCF detection is 7.3$\sigma$, 
indicating that, as expected, $^{13}$C$^{16}$O is the second most common form of CO.
Although the isotopologue detections are admittedly marginal, these results suggest that with future higher signal-to-noise NIRSpec data 
 it will be possible to search for 
  individual isotopologues of CO or other molecules by looking at how the significance of CCF detections increases when adding one isotopologue at a time.

\section{Acknowledgments}
\begin{acknowledgments}
This work is based on observations made with the NASA/ESA/CSA JWST. The data were obtained from the Mikulski Archive for Space Telescopes at the Space Telescope Science Institute, which is operated by the Association of Universities for Research in Astronomy, Inc., under NASA contract NAS 5-03127 for JWST. The specific observations can be accessed via \dataset[doi:10.17909/gdm2-0q65]{http://dx.doi.org/10.17909/gdm2-0q65}. These observations are associated with program JWST-ERS-01366. Support for program JWST-ERS-01366 was provided by NASA through a grant from the Space Telescope Science Institute. This work was supported by grant JWST-ERS-01366.033-A. E.E-B and E.P. acknowledge funding from the Spanish Ministry of Economics and Competitiveness through project PGC2018-098153-B-C31. E.E-B acknowledges financial support from the European Union and the State Agency of Investigation of the Spanish Ministry of Science and Innovation (MICINN) under the grant PRE2020-093107 of the Pre-Doc Program for the Training of Doctors (FPI-SO) through FSE funds. J.K. acknowledges financial support from Imperial College London through an Imperial College Research Grant.
\end{acknowledgments}

\newpage
\vspace{5mm}
\facilities{JWST(NIRSpec)}


\software{
\texttt{batman} \citep{kreidberg2015},
\texttt{emcee} \citep{emcee},
\texttt{ExoTIC-LD}
\citep{ExoticLD},
\texttt{george} \citep{ambikasaran2015},
\texttt{petitRADTRANS} \citep{Petitradtrans},
\texttt{PyAstronomy} \citep[][\url{https://github.com/sczesla/PyAstronomy}]{pyastronomy},
\texttt{spectres} \citep{spectres},
\texttt{scipy} \citep{scipy},
\texttt{Tiberius} \citep{2017MNRAS.468.3907K,Kirk18,Kirk19,Kirk21}
}




\bibliography{sample631}{}
\bibliographystyle{aasjournal}



\clearpage
\newpage
\appendix

\section{White and Spectroscopic light curves adopted parameters and priors}\label{app: pars}
    
\setcounter{table}{0}
\renewcommand{\thetable}{A.\arabic{table}}

\begin{table*}[!ht]
\tablewidth{\textwidth}
  \caption{Adopted parameters for the white light curve (WLC) fitting, including whether the value was fixed in the fitting or if a uniform ($\cal{U}$) or Gaussian ($\cal{G}$) prior was used instead. For uniform priors, the lower and upper bounds are given in brackets. For Gaussian priors, the mean and standard deviation of the priors are given in brackets. We also list the posterior distribution of each parameter after the fitting. The posteriors of the WLC fitting were used as starting values for the spectroscopic light curves (SLC) fitting.}\label{tab:adopted_parameters}
  \centering
  
  
      \begin{tabular}{lcccc}
      \toprule \toprule
      
      \bf Parameter & WLC starting value$^{(1)}$ & WLC Prior type & WLC Posterior / SLC starting value & SLC Prior type  \\ 
      \midrule
      
      \textit{P} [days] &  4.0552941 (34) & Fixed & -- & Fixed \\
      \textit{e} & 0 & Fixed & -- & Fixed \\
      \textit{$\rm T_{0}$} [BJD] & 2 455 342.96913 (63) & $\cal{U}$($T_0-2$, $T_0+2$) &2459791.61206 (89) & Fixed \\
      \textit{i} [$\rm \deg$] & $87.32$ & $\cal{U}$(80,90) &  $87.81^{+0.20}_{ -0.18}$ & Fixed \\
      \textit{a/$R_{s}$} & $11.55$ & $\cal{G}$(11.55,0.17) & $11.425\pm0.015$ & Fixed \\
      \textit{$\rm R_{p}/R_{s}$} & 0.14052 & $\cal{U}$(0., 0.3) & 0.14636 $^{+0.00068}_{-0.00071}$ & $\cal{U}$(0.,0.3) \\
      \bottomrule \bottomrule
      \multicolumn{5}{l}{$^{(1)}$ Values adopted from \cite{Mancini18}.}
          
      \end{tabular}
\end{table*}


\newpage
\section{Spectrum Sampling Tests}\label{app: sampling}

To determine whether to use the observed pixel-column sampled or Nyquist limit sampled spectrum of WASP-39b as input to our CCF analysis in section~\ref{sec:CCSearch}, we followed the same procedure described in that section, but replacing the observed spectrum of WASP-39b by synthetic spectra.


We generated synthetic CO spectra of WASP-39b following similar steps to the ones described in section~\ref{sec:COtemplate}.  
We generated a CO \textit{all isotopologues} transmission spectrum of WASP-39b with \texttt{petitRADTRANS} \citep{Petitradtrans} at resolving power R=$10^{6}$ following as described in section~\ref{sec:COtemplate}, and 
then lowered the resolution of the spectrum to match the 1-pixel resolution of our NIRSpec G395H NRS2 observations using the \texttt{spectres} module.
From that spectrum we produced four different synthetic spectra:
1-pixel sampled; 
2-pixel sampled starting on the first pixel of the array, hereafter\textit{ 2-pixel (0)}, 2-pixel sampled starting on the second pixel of the array,  hereafter\textit{ 2-pixel (1)},  and 3-pixel sampled.
We normalized each of the synthetic spectra using a fourth order polynomial, as in section~\ref{sec:COtemplate}, then selected the portion of the spectrum between 4.6 and 5.0 $\mu$m used in our CO search,  and added noise using the distribution of noise in the observed spectrum between those wavelengths.
As a last step 
we offsetted each synthetic spectrum  in wavelength by -87.33 ${\rm km s^{-1}}$ to match the expected systemic velocity. 

\begin{figure*}[ht!]
\gridline{\fig{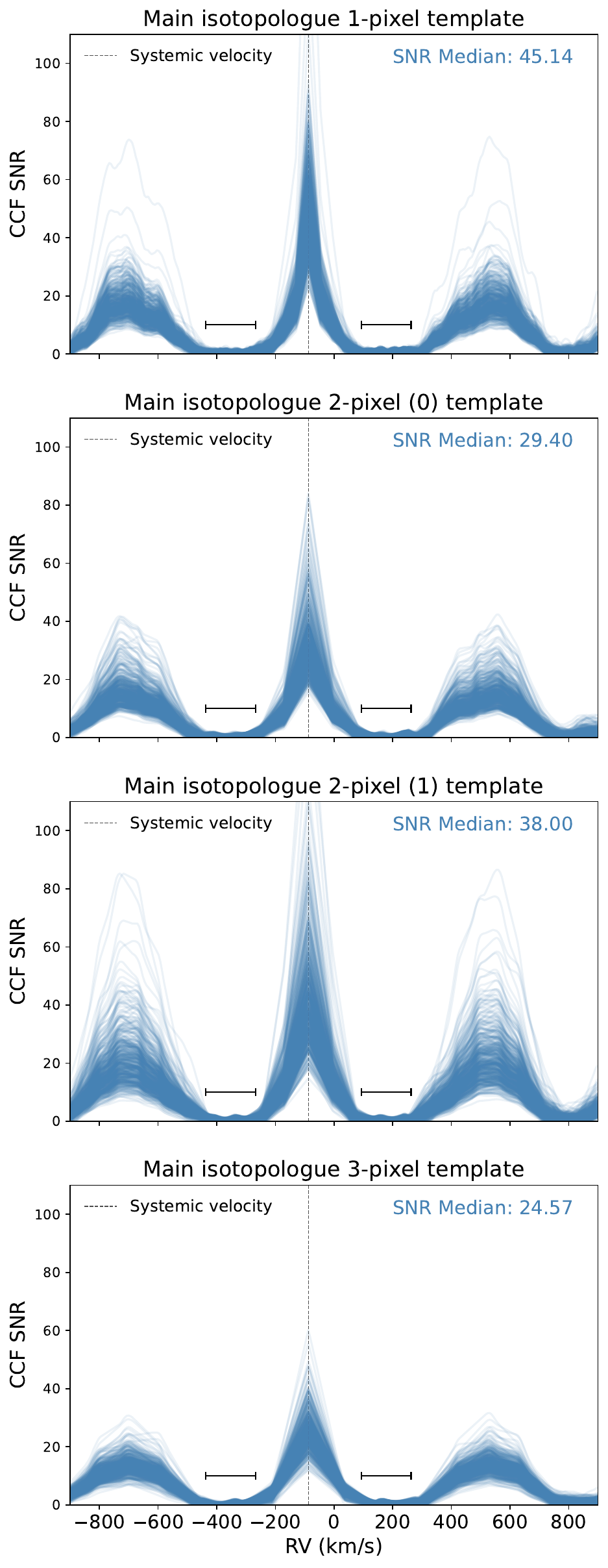}{0.45\columnwidth}{}\fig{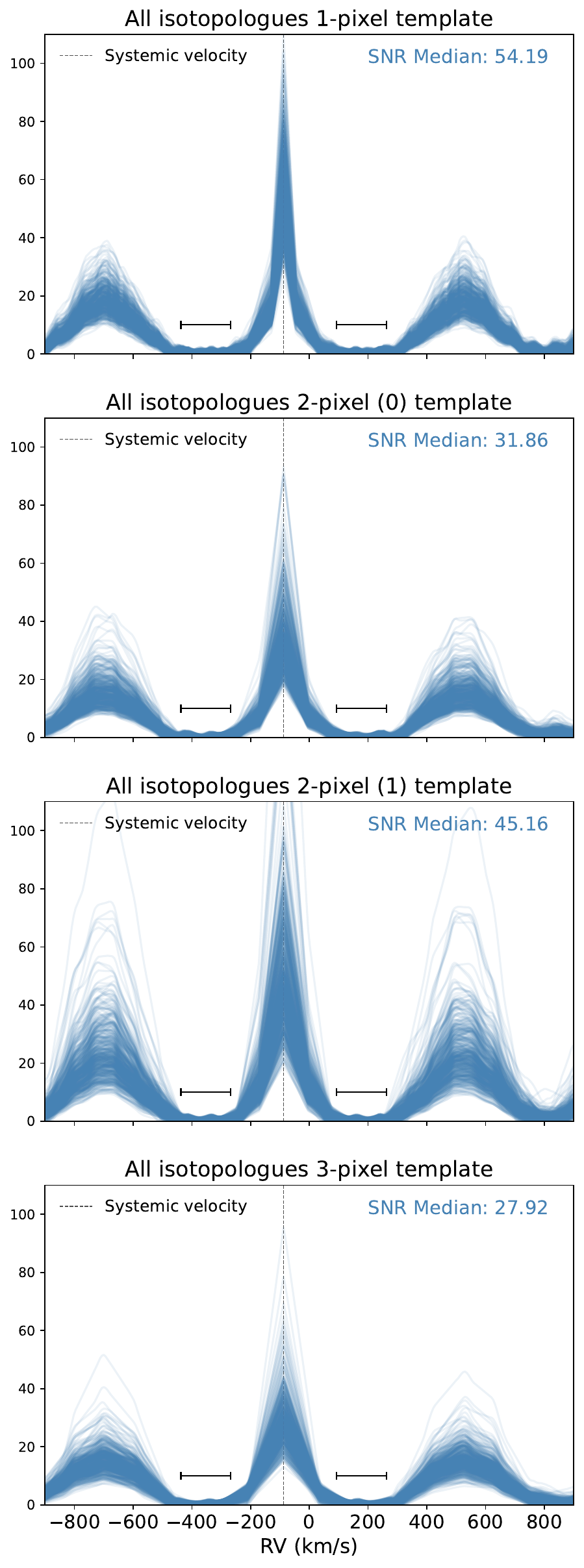}{0.432\columnwidth}{}}
\caption{Cross-correlation functions (CCFs) resulting from the cross-correlation of the synthetic transmission spectra with the CO main isotopologue (left panels) and the CO all isotopologues templates (right panels) in the 4.6--5.0\,$\rm \mu$m range using a radial velocity step of 6~km/s. The first row panels show the resulting CCFs for the 1-pixel sampled transmission spectrum. The second and third rows panels show the resulting CCFs for the 2-pixel sampled transmission spectra, starting in the first and second pixel, respectively. The third row panels show the resulting CCFs for the 3-pixel sampled transmission spectrum. The black horizontal bars indicate the continuum regions used to calculate the signal-to-noise ratio of the CCF, and calculate the significance of the central peak. The black-dashed line shows the systemic velocity relative to the JWST at the time of the observation. Additionally, the median signal-to-noise ratio value of the main peak is printed in each panel.
\label{fig:appendix_synth}}
\end{figure*}

We cross-correlated each synthetic spectrum with the CO \textit{main-isotopologue} and CO \textit{all-isotopologues} templates described in Section~\ref{sec:COtemplate}, following the same procedure as with the real data described in section~\ref{sec:CCSearch}. The results are summarized in Figure~\ref{fig:appendix_synth}. The figure shows that the CO signal is detected in all cases and with a CCF shape that is very similar to the one produced by the real data, including a main peak centered at the systemic velocity, and secondary peaks at a mean distance of $\pm$ 624 ${\rm km s^{-1}}$ from the central peak.  In addition, the SNRs of the central peaks  are consistently larger in the cases when we use the \textit{all isotopologues} template, confirming the increase in signal that we see in the real data when we use a template with multiple isotopologues.

The simulations show that 1-pixel sampling produces the same CCF results, but with higher SNR than the other samplings. Therefore we decided to use 1-pixel sampling in our analysis. We note that the peak SNRs of the CCFs in the simulation are significantly larger that the values obtained in the real data, but we attribute that difference to the simulated spectra only containing CO, which in the real data, there are also contributions from ${\rm H_2O}$, ${\rm CO_2}$ lines, stellar  ${\rm CO}$ lines \citep{Deming2017}, in that wavelength region that can dilute the CO signal.







\newpage
\section{$\rm^{12}C^{16}O$ + $\rm^{13}C^{16}O$ cross-correlation results}\label{app: 13C16O}

\begin{figure}[ht!]
\epsscale{0.5}
\plotone{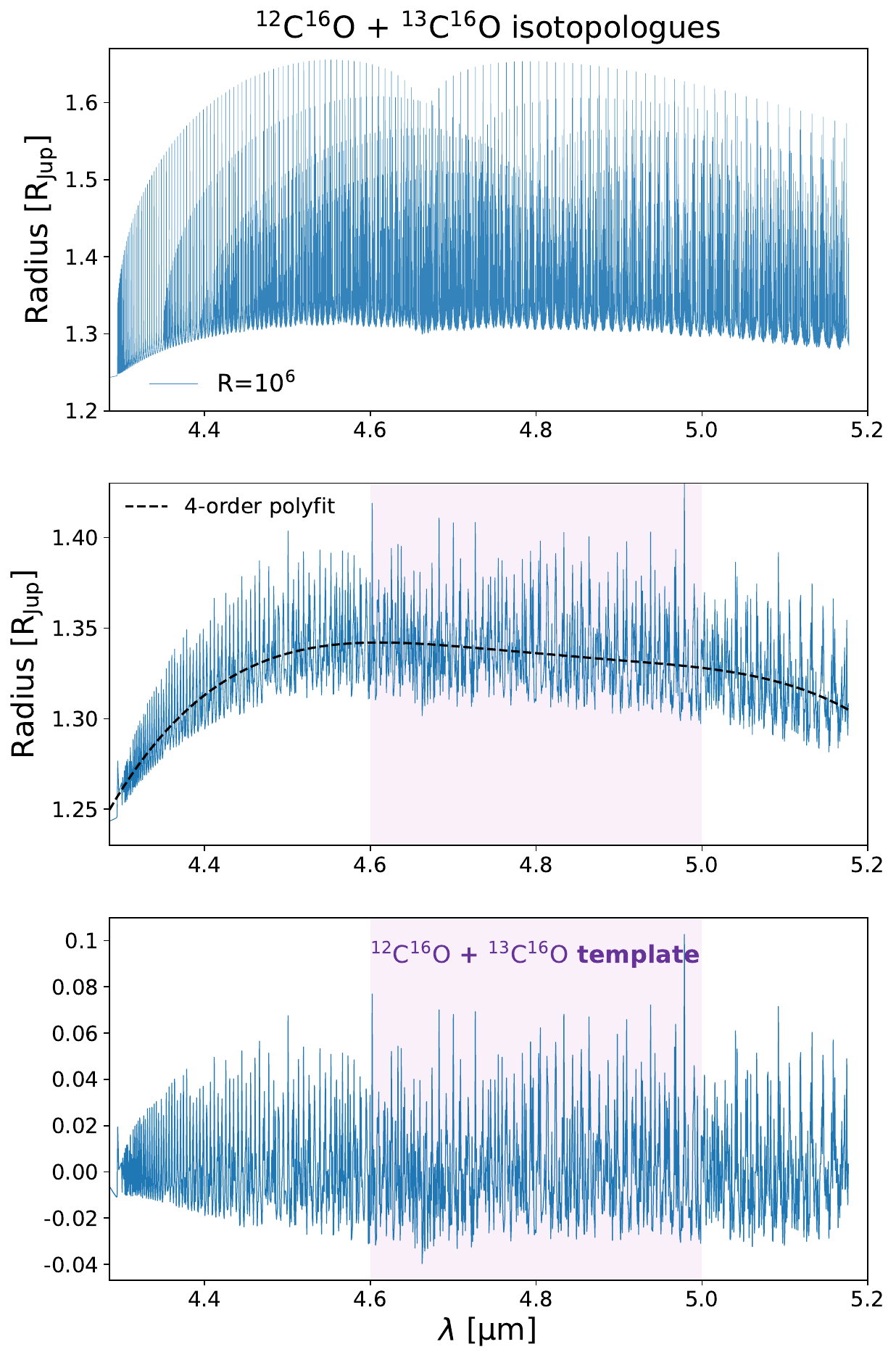}
\caption{Atmospheric models for the combination of $\rm^{12}C^{16}O$ ($\rm 98.7\%$) and $\rm^{13}C^{16}O$ ($\rm 1.3\%$). Top: High-resolution ($10^{6}$) \texttt{petitRADTRANS}-computed transmission spectrum model. Middle: CO spectrum model degraded to match the sampling of the observed transmission spectrum. The region used for the cross-correlation search is highlighted in purple. The black dashed line shows the 4-order polynomial fit that is used for the normalization. Bottom: Normalized CO template used in the cross-correlation analysis.
\label{fig:13C16O_template}}
\end{figure}

\begin{figure}[ht!]
\gridline{\fig{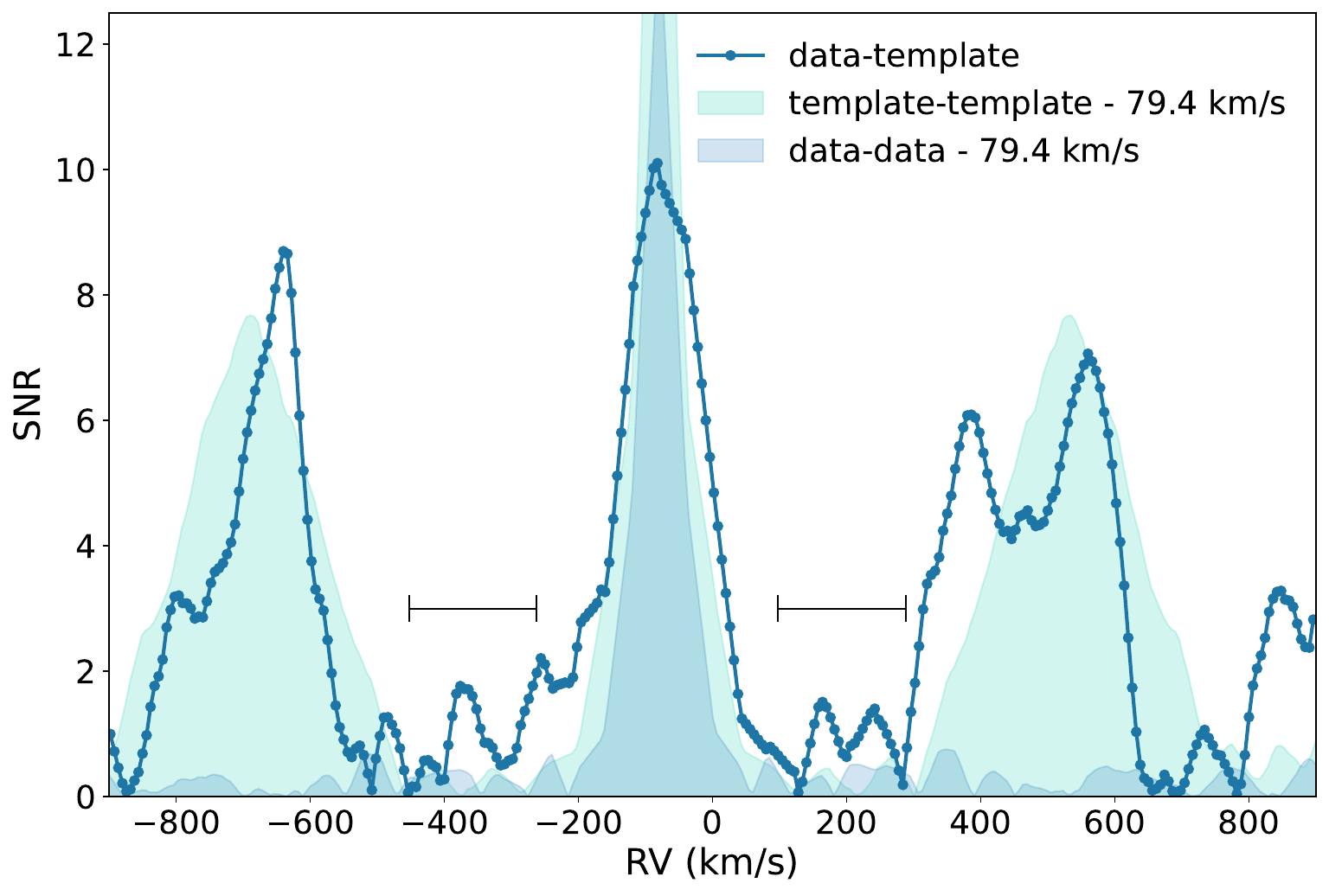}{0.6\columnwidth}{(a)}
\fig{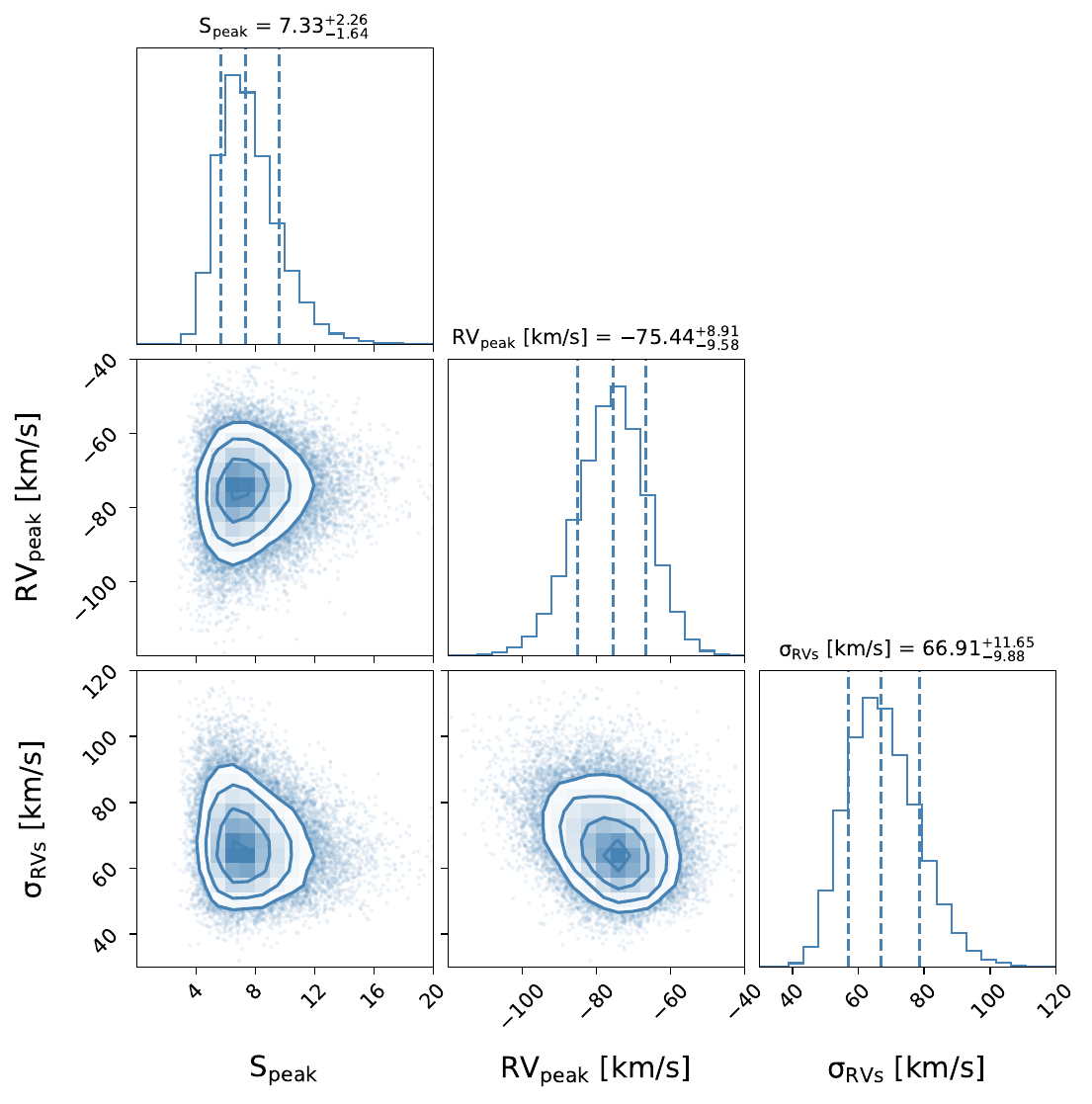}{0.4\columnwidth}{(b)}}
\caption{(a) Cross-correlation function (CCF) of the WASP39b transmission spectrum with
the $\rm^{12}C^{16}O$ ($\rm 98.7\%$) + $\rm^{13}C^{16}O$ ($\rm 1.3\%$) template. We compare the CCFs resulting from the self cross-correlation of the template (light blue area) and the spectrum (dark blue area), which were scaled arbitrarily. (b) Distribution of the best-fit Gaussian parameters (significance of the central peak, ${\rm S_{peak}}$, the central peak velocity, ${\rm RV_{peak}}$, and the FWHM of the distribution of velocities, ${\rm \sigma_{RVs}}$) retrieved from the sampled transmission spectra that were randomly generated following a Gaussian distribution centered in the observed transmission spectrum and considering the errors as the Gaussian standard deviation in each point. We performed 30000 iterations. 
\label{fig:13C16O_CC}}
\end{figure}

We performed an additional cross-correlation analysis of the observed transmission spectrum using a template composed of the two most abundant CO isotopologues: $\rm^{12}C^{16}O$ and $\rm^{13}C^{16}O$. We generated the template following the procedure detailed in Section~\ref{sec:COtemplate}, setting the relative abundances of the isotopologues to match those of 
Earth,  
i.e., $\rm 98.7\%$  $\rm^{12}C^{16}O$ and  $\rm 1.3\%$ $\rm^{13}C^{16}O$. Figure~\ref{fig:13C16O_template} shows the model spectrum used as template in this case. The CCF results, shown in Figure~\ref{fig:13C16O_CC}), are similar to the ones obtained using the CO \textit{all-isotopologues} template. The significance of the detection, being $\rm 7.33^{+2.26}_{-1.04}$ sigma, increases compared to the
 CO \textit{main-isotopologues} result but is slightly below the \textit{all-isotopologues} significance, indicating that the \textit{all-isotopologues} template produces a better match.

\end{document}